\documentclass[10pt,twocolumn]{article}
\usepackage[a4paper,margin=0.75in,columnsep=0.25in]{geometry}
\usepackage[T1]{fontenc}
\usepackage{newtxtext,newtxmath}
\usepackage{microtype}
\usepackage{graphicx}
\usepackage{booktabs}
\usepackage{array}
\usepackage{multirow}
\usepackage[table]{xcolor}
\usepackage{amsmath}
\usepackage{tikz}
\usetikzlibrary{arrows.meta,positioning,fit,calc}
\usepackage[hidelinks]{hyperref}
\usepackage[capitalise,noabbrev]{cleveref}
\usepackage{caption}
\usepackage{enumitem}
\setlist{nosep,leftmargin=*}
\usepackage{natbib}
\setcitestyle{numbers,square}

\definecolor{ink}{HTML}{0B0B0B}
\definecolor{ink2}{HTML}{52514E}
\definecolor{sblue}{HTML}{2A78D6}
\definecolor{rowhl}{HTML}{EAF2FC}
\newcommand{\bthree}{B$^3$-F1}

\title{\textbf{VoicePalette:}\\[2pt]
\textbf{Song-Context Singer Embeddings for Line-Level Singer Diarization}}
\author{Allen Ng\\[2pt]
\normalsize nagiflow LLC, Tokyo, Japan\\
\normalsize\texttt{allen.ng@nagiflow.com}}
\date{}

\begin{document}
\maketitle

\begin{abstract}
Line-level singer diarization assigns each lyric line of a song to the member who sings it. The strongest published
pipeline embeds each line separately with a singing-adapted ECAPA2 encoder and only then reasons over the song, with
a sticky Viterbi pass over line order and linking of a group's songs. VoicePalette moves song context into the model: a
light attention head, initialized to reproduce the per-line voiceprints exactly, re-embeds every line in the light of
the whole song, on top of a fine-tuned ReDimNet-M encoder and an input that averages the lead and all-vocal stems; a
partition is chosen among k-means candidates by its own objective. VoicePalette is a single model of one encoder
and one head, evaluated once as pre-registered within a budget of six test looks. On 168 songs of 13 artists unseen in
training it improves \bthree{} from 0.522 to 0.555 over that pipeline ($+0.033$, 95\% CI $[+0.022, +0.044]$,
$p<0.001$, significant after correction for all six looks) and accuracy from 54.2\% to 58.8\%; on the verified
held-out songs \bthree{} rises from 0.544 to 0.601. It matches an ensemble of five song-context heads on \bthree{} and exceeds it in
accuracy. On ColorCoded-SD and IdolSongsJp the gains ($+0.021$, $+0.015$) are within noise. We also report thirteen
ablations that tie or lose and an oracle analysis: the right partition is among the candidates for 72.8\% of
development lines but is chosen for 63.9\%, and lines sung far from the singer's usual register remain the main
error.
\end{abstract}

% ------------------------------------------------------------------------------------------------
\section{Introduction}
\label{sec:intro}

Line-level singer diarization groups the lyric lines of a song by the member who sings them~\citep{ng2026linelevel}.
In multi-member pop groups the members are chosen to blend, trade lines every few seconds and are layered with each
other's harmonies, so a voiceprint computed for each line in isolation confuses them. The best published system~\citep{ng2026linelevel}
reaches 0.699 \bthree{} on the ColorCoded-SD benchmark by isolating the lead vocal with a karaoke separator,
embedding each line with ECAPA2 fine-tuned on singing (ECAPA2-ft), and decoding each song with two training-free
steps: a sticky Viterbi pass that favours keeping the singer of the previous line, and discography linking, which
matches clusters across a group's songs. All of its song-level reasoning sits after the embedding: the line
voiceprints themselves never see the rest of the song.

This paper moves song context into the model. A light attention head reads all lines of a song at once and re-embeds
each line so that lines of the same singer move together; it is initialized to reproduce the per-line voiceprints
exactly and trained with an in-song contrastive loss. On top of it, a fine-tuned ReDimNet-M encoder, an input that
averages the lead and all-vocal stems, and a partition chosen among k-means candidates by its own objective give VoicePalette (a song's singers form a palette of voices, and each line gets one colour), a
\emph{single} model (one encoder and one head, no ensembling) that we evaluate once, as pre-registered, on four
test sets.

The second theme is method. This work was developed under a fixed budget of six looks at the test sets, with every
decision taken on an artist-disjoint development set whose labels two annotators confirm. That protocol changed what
we believe: corpus validation repeatedly preferred changes that lost on the development set, five-seed means moved by
half a point between seed sets, and of the many ideas that looked promising on one seed or one validation split, none
survived ten seeds and a paired interval. We report these negative results with the same statistics as the positive
ones, because they bound where the remaining error is.

\paragraph{Contributions.}
\begin{itemize}
  \item \textbf{A song-context head} with a computed initialization (step 0 equals the per-line voiceprints, so
  training can only move away from a known pipeline) and training choices for sparse songs and lead-track input
  (\cref{sec:method}).
  \item \textbf{VoicePalette, a single-model system} (encoder + head + candidate selection) evaluated once on ColorCoded-SD,
  IdolSongsJp, the held-out verified songs and 168 songs of 13 held-out artists, against the ECAPA2-ft pipeline and a
  five-head ensemble (\cref{sec:results}).
  \item \textbf{An evaluation protocol} for small, noisy music benchmarks: an artist-disjoint development set with
  cross-checked labels, ten-seed paired intervals, and a logged test-look budget with multiple-comparison correction
  (\cref{sec:protocol}).
  \item \textbf{Bounds and negative results:} the right partition is among the candidates far more often than it is
  chosen (72.8\% vs.\ 63.9\% on dev), and thirteen changes to encoder, head, input and decoding are ties or losses
  (\cref{sec:negative}).
\end{itemize}

% ------------------------------------------------------------------------------------------------
\section{Related Work}
\label{sec:related}

\paragraph{Diarization with clustering and context.} Speaker diarization combines embeddings with clustering, often
refined over time~\citep{landini2022vbx,park2022review}; end-to-end models with vector clustering combine local neural
diarization with global clustering~\citep{kinoshita2021eendvc}. Short segments carry unreliable embeddings, which
uncertainty-aware clustering down-weights~\citep{silnova2020probabilistic}. Our head is a small
transformer~\citep{vaswani2017attention} over the \emph{lines} of one song rather than over frames; clustering
remains a separate, label-free step.

\paragraph{Singing voices.} Speaker encoders degrade on singing and mixtures~\citep{suda2024fruitsmusic}; ECAPA2-ft~\citep{ng2026linelevel}
adapts ECAPA2~\citep{thienpondt2023ecapa2} to singing with in-song contrastive negatives~\citep{khosla2020supcon}. Here we adapt
ReDimNet~\citep{yakovlev2024redimnet}, a speaker network that reshapes between 2D and 1D feature maps. Self-supervised
singer representations~\citep{torres2023singer} did not help in our fusion experiments (\cref{sec:negative}).

\paragraph{Separation.} We use HT-Demucs~\citep{rouard2023htdemucs} vocals and a Mel-RoFormer karaoke
model~\citep{becruily2025karaoke} for the lead vocal, as in~\citep{ng2026linelevel}, and test BS-RoFormer~\citep{lu2024bsroformer} as a
further view.

% ------------------------------------------------------------------------------------------------
\section{Task, Data and Protocol}
\label{sec:protocol}

\paragraph{Task.} As in~\citep{ng2026linelevel}: given a song and the start and end times of its single-singer lines, partition the
lines into the song's $k$ singers; $k$ is known from the member list. We score micro \bthree{}~\citep{bagga1998bcubed,
amigo2009comparison} over lines and one-to-one line accuracy after Hungarian mapping~\citep{kuhn1955hungarian}
(``Acc.''), with the ColorCoded-SD scoring tools~\citep{ng2026linelevel}.

\paragraph{Training corpus.} Line-coloring projects from the ColorCoded application (2,277 songs, single unverified
annotations). Songs of all test artists and of the 21 development artists are excluded, as are
songs whose step-0 leave-one-out centroid accuracy is below 0.4 (mostly songs with many singers and noisy
labels). The song heads train on 1,079 songs and select checkpoints on 111 more.

\paragraph{Development set.} 27 songs by 21 artists (1,336 lines, median five singers) whose labels two different
users agree on ($\ge$75\% of time), with audio taken from the best-agreeing project. Its artists are excluded from
every encoder and head, so it measures unseen singers, like the tests. All model decisions in this paper were taken
on it (with corpus validation where stated); with 27 songs it resolves differences of about 1.5 accuracy points.
One song has no lead track and uses Demucs vocals only.

\paragraph{Test sets.} \emph{ColorCoded-SD} (43 songs, 1,740 lines) and the 14 labelled \emph{IdolSongsJp}
songs~\citep{suda2025idolsongsjp} (309 lines, J-pop session singers) are the public benchmarks of~\citep{ng2026linelevel}.
Thirteen further artists were removed from all training data before any encoder in this line of work was trained.
\emph{Held-out (verified)} are 19 two-annotator songs of eight of them.
\emph{Held-out artists} are the 168 other corpus songs of these 13 artists (7,809 lines, 14 artist keys,
single-annotator labels; fan covers and the 19 verified songs removed), which no encoder, head or decision has seen.
ColorCoded-SD is scored on the annotators' original line timings; with the released timings, which lie on a 0.25\,s
grid, VoicePalette scores 0.722 \bthree{} and 74.7\% accuracy on our copies of the songs.

\paragraph{Test-look budget.} We looked at the test sets six times, each look pre-registered in the research log
before scoring: (1) a bandmate-interference encoder (tie), (2) window voiceprints (not better), (3) lead-track views
for the head (seed spread as large as the effect), (4) the five-head ensemble on the first three test sets, (5) the
same ensemble on the 168 held-out-artist songs, and (6) VoicePalette. We correct the primary
comparison of look~6 for all six looks~\citep{holm1979simple}. Intervals are 95\% percentiles of 2,000 paired
bootstrap resamples of songs (tests) or of artists (dev); $p$-values are two-sided.

\paragraph{Reference systems.} \emph{ECAPA2-ft + sticky + linking}~\citep{ng2026linelevel} embeds each line
(lead input) with ECAPA2-ft and decodes it in two training-free steps. \emph{Sticky decoding} starts from AHC and
alternates singer centroids with a Viterbi pass that maximizes the cosine of each line to its centroid plus a bonus
$\beta=0.2$ for keeping the previous line's singer. \emph{Discography linking} clusters every song of a group, links
each song's centroids one-to-one (Hungarian) to the group's members, re-estimates a prototype per member, and
re-decodes each song with the sticky pass against a blend of member prototype and song centroid.
On ColorCoded-SD we use its published predictions; on the other sets we rerun it with an ECAPA2-ft trained without
the held-out artists. The \emph{five-head ensemble} is an earlier system of this work: five song-context heads
(\cref{sec:head}, different seeds) on ECAPA2-ft embeddings, outputs averaged, decoded by average-linkage AHC.

\paragraph{Lead protocol.} A line's lead view is used when the karaoke track is not near-silent on it (line RMS at
least $0.25\times$ the song's mean lead-line RMS), as in~\citep{ng2026linelevel}; otherwise the line uses Demucs vocals only.

% ------------------------------------------------------------------------------------------------
\section{Method}
\label{sec:method}

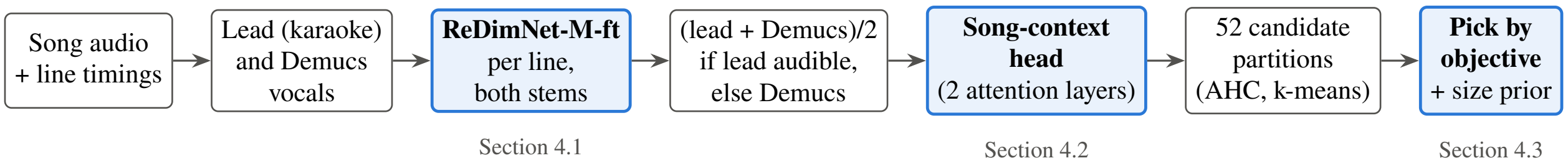
\begin{figure*}[t]
\centering
\begin{tikzpicture}[
  font=\footnotesize, node distance=4mm,
  box/.style={draw=ink2, line width=0.5pt, rounded corners=2pt, fill=white, align=center, minimum height=10mm, inner sep=3pt},
  hl/.style={box, draw=sblue, line width=0.9pt, fill=rowhl},
  arr/.style={-{Stealth[length=2mm]}, line width=0.6pt, color=ink2}]
\node[box] (song) {Song audio\\+ line timings};
\node[box, right=of song] (sep) {Lead (karaoke)\\and Demucs\\vocals};
\node[hl, right=of sep] (enc) {\textbf{ReDimNet-M-ft}\\per line,\\both stems};
\node[box, right=of enc] (avg) {(lead + Demucs)/2\\if lead audible,\\else Demucs};
\node[hl, right=of avg] (head) {\textbf{Song-context}\\\textbf{head}\\(2 attention layers)};
\node[box, right=of head] (cand) {52 candidate\\partitions\\(AHC, k-means)};
\node[hl, right=of cand] (sel) {\textbf{Pick by}\\\textbf{objective}\\+ size prior};
\foreach \a/\b in {song/sep, sep/enc, enc/avg, avg/head, head/cand, cand/sel} \draw[arr] (\a) -- (\b);
\node[below=1.5mm of enc, text=ink2, font=\scriptsize] {\cref{sec:encoder}};
\node[below=1.5mm of head, text=ink2, font=\scriptsize] {\cref{sec:head}};
\node[below=1.5mm of sel, text=ink2, font=\scriptsize] {\cref{sec:decode}};
\end{tikzpicture}
\caption{VoicePalette. One encoder and one head; separation is a fixed front end. Highlighted: the parts that
changed the result on the development set (\cref{tab:ladder}).}
\label{fig:pipeline}
\end{figure*}

\subsection{Encoder}
\label{sec:encoder}
Each line (its first 15\,s, padded to at least 0.2\,s) is embedded directly from the 16\,kHz waveform by
ReDimNet-M~\citep{yakovlev2024redimnet} (4.8M parameters, VoxCeleb-trained), fine-tuned as ECAPA2-ft~\citep{ng2026linelevel}: supervised
contrastive loss with in-song negatives weighted $\times2$, temperature 0.1, and an anchor $0.2\,(1-\cos(z,z_0))$ to
the frozen speech embedding $z_0$; AdamW, learning rate $5\cdot10^{-5}$, 500 steps, BatchNorm frozen. The stem and
the first five stages stay frozen; stage~6, the feature aggregation and the pooling and projection layers (2.2M
parameters) are trained. Fine-tuning takes 17\,min on the GPU of a Mac mini. Frozen, ReDimNet-M was below a frozen
ECAPA2 on our data (58.6\% vs.\ 60.7\% dev accuracy); fine-tuned, it is the largest single gain of this paper
(\cref{tab:ladder}).

\subsection{Song-context head}
\label{sec:head}
A song's lines in time order, with embeddings $x_1,\dots,x_N$, are centred on the song mean and $\ell_2$-normalized,
then passed through two layers of four-head self-attention with a feed-forward block and residual connections:
\begin{equation}
x \leftarrow x + O\,\mathrm{Attn}(Qx, Kx, Vx;\,b\,A),\qquad x \leftarrow x + \mathrm{FF}(x),
\end{equation}
where $A$ marks adjacent lines and $b$ is a learned turn bias. The output is re-normalized. The initialization is
\emph{computed}: $Q=K=\tau^{-1/2}I$ with $\tau=0.2$, so each line attends to the lines that already sound like it;
$V=I$; $b=\log\big(P(\text{same}\mid\text{adjacent})/P(\text{same})\big)=0.97$ from annotation statistics; and the
output projection $O$ and the last feed-forward layer start at zero. Step 0 therefore reproduces the per-line
voiceprints exactly, while gradients already flow through a meaningful attention pattern. The head has 0.6M
parameters.

\paragraph{Training.} Each step draws eight training songs; the loss is the supervised contrastive
loss~\citep{khosla2020supcon} over the song's lines (temperature 0.1). Three choices matter. \emph{Variable line
dropout} removes a fraction $d\sim U(0,0.7)$ of a song's lines, so the head also sees sparse songs; without it the head
over-relies on turn-taking and loses on IdolSongsJp, whose songs have 3.6 lines per singer against 7.8. \emph{Lead
views:} for songs with a karaoke track, each line uses the lead embedding with probability 0.5, matching the test
input. \emph{Crops:} with probability 0.3 a line is replaced by a 0.2--1.5\,s crop, so short lines learn to borrow
evidence from the song. Training runs 6,000 steps on a CPU in about a minute, and the checkpoint with the best
validation \bthree{} is kept.

\subsection{Input and decoding}
\label{sec:decode}
At test time each line is the average of its lead and Demucs embeddings when the lead is audible, and Demucs
otherwise. The averaged view was found on the development set and is better than either stem alone
(\cref{tab:ladder}). With the true $k$ we form 52 candidate partitions of the head's output $z$: average-linkage AHC,
spherical k-means started from it, and 50 k-means++~\citep{arthur2007kmeanspp} restarts with a seed derived from the
song. We pick
\begin{equation}
\hat P=\arg\max_P\ \frac{1}{N}\sum_{j=1}^{k}\Big\lVert\sum_{\ell\in P_j} z_\ell\Big\rVert+\lambda\,H(P),
\end{equation}
the spherical k-means objective plus a small entropy bonus on the cluster sizes ($\lambda=0.01$, chosen on corpus
validation). The bonus counters the objective's preference for absorbing a minor member into a larger cluster.

% ------------------------------------------------------------------------------------------------
\section{Results}
\label{sec:results}

\subsection{Test results}
\label{sec:test}

\begin{table*}[t]
\centering\small
\caption{Test look~6 (pre-registered). The reference systems (\cref{sec:protocol}) are scored on the same songs and
lines; VoicePalette is the mean of ten head seeds (range of accuracy in parentheses), each one model. $\Delta$:
VoicePalette minus ECAPA2-ft + sticky + linking, paired song bootstrap (2,000), two-sided $p$; $^{\ast}$\,$p<0.05$.
$^{\dagger}$Primary comparison: 168 songs of the 13 held-out artists, single-annotator labels, lead protocol
(\cref{sec:protocol}); its $p$ is below $0.05/6$, the Bonferroni level for six looks.}
\label{tab:test}
\begin{tabular}{@{}llccrcc@{}}
\toprule
Test & System & \bthree{} & Acc.\ (\%) & $\Delta$\bthree{} & 95\% CI & $p$ \\
\midrule
\multirow{3}{*}{\shortstack[l]{ColorCoded-SD\\{\footnotesize 43 songs, 1,740 lines}}} & ECAPA2-ft + sticky + linking~\citep{ng2026linelevel} & 0.699 & 72.4 & --- & & \\
 & Five-head ensemble (ECAPA2-ft, AHC) & 0.720 & 72.6 & --- & & \\
 & \textbf{VoicePalette} (10-seed mean) & \textbf{0.720} & \textbf{74.7} {\footnotesize (73.8--75.5)} & +0.021 & [$-$0.015, +0.056] & 0.252 \\
\midrule
\multirow{3}{*}{\shortstack[l]{IdolSongsJp\\{\footnotesize 14 songs, 309 lines}}} & ECAPA2-ft + sticky + linking~\citep{ng2026linelevel} & 0.866 & 85.1 & --- & & \\
 & Five-head ensemble (ECAPA2-ft, AHC) & 0.861 & 84.1 & --- & & \\
 & \textbf{VoicePalette} (10-seed mean) & \textbf{0.881} & \textbf{88.1} {\footnotesize (85.4--89.3)} & +0.015 & [$-$0.021, +0.050] & 0.428 \\
\midrule
\multirow{3}{*}{\shortstack[l]{Held-out (verified)\\{\footnotesize 19 songs, 903 lines}}} & ECAPA2-ft + sticky + linking~\citep{ng2026linelevel} & 0.544 & 55.9 & --- & & \\
 & Five-head ensemble (ECAPA2-ft, AHC) & 0.575 & 57.7 & --- & & \\
 & \textbf{VoicePalette} (10-seed mean) & \textbf{0.601} & \textbf{62.4} {\footnotesize (61.1--63.7)} & +0.058$^{\ast}$ & [+0.028, +0.085] & $<$0.001 \\
\midrule
\multirow{3}{*}{\shortstack[l]{Held-out artists$^{\dagger}$\\{\footnotesize 168 songs, 7,809 lines}}} & ECAPA2-ft + sticky + linking~\citep{ng2026linelevel} & 0.522 & 54.2 & --- & & \\
 & Five-head ensemble (ECAPA2-ft, AHC) & 0.558 & 56.8 & --- & & \\
 & \textbf{VoicePalette} (10-seed mean) & \textbf{0.555} & \textbf{58.8} {\footnotesize (58.5--59.1)} & +0.033$^{\ast}$ & [+0.022, +0.044] & $<$0.001 \\
\bottomrule
\end{tabular}

\end{table*}

\Cref{tab:test} gives the single test look. On the primary comparison, the 168 songs of artists that no part of the
system has seen, VoicePalette improves \bthree{} over ECAPA2-ft + sticky + linking by $+0.033$ ($[+0.022, +0.044]$, $p<0.001$) and
accuracy by 4.6 points (54.2\% to 58.8\%); every one of the ten seeds is above it (\bthree{} 0.552--0.559). The
gain survives correction for all six test looks. On the 19 verified held-out songs the improvement is larger
($+0.058$ \bthree{}, $+6.5$ accuracy points, $p<0.001$). On ColorCoded-SD ($+0.021$) and IdolSongsJp ($+0.015$) the
direction is the same but the intervals include zero; both sets are small (43 and 14 songs).

Against the five-head ensemble (secondary, uncorrected), VoicePalette ties on \bthree{} (168 songs: $-0.003$,
$[-0.013, +0.007]$, $p=0.57$; ColorCoded-SD $+0.001$) and is more accurate on all four tests ($+2.0$, $+2.1$, $+4.0$
and $+4.7$ points; significant on the 168 songs, IdolSongsJp and held-out). One head therefore replaces the five-head
ensemble. The development target of 65\% accuracy was not reached: VoicePalette
averages 63.9\% on development (best seed 65.3\%), so we report it, not claim it.

\subsection{How VoicePalette was built}
\label{sec:ladder}

\begin{table}[t]
\centering\small
\caption{Development set (27 songs, 21 unseen artists, 1,336 lines): line accuracy (\%), mean over song-head seeds.
Each row adds one change. The first two rows are the reference systems of \cref{sec:protocol}.}
\label{tab:ladder}
\setlength{\tabcolsep}{3pt}
\begin{tabular}{@{}lrr@{}}
\toprule
System & Seeds & Acc. \\
\midrule
ECAPA2-ft + sticky + linking$^a$ & --- & 60.0 \\
Five-head ensemble (ECAPA2-ft, AHC) & 5 & 59.5 \\
\midrule
ECAPA2-ft + one head, Demucs, AHC & 5 & 59.3 \\
\quad + (lead + Demucs)/2 input & 5 & 60.1 \\
\quad + pick by k-means objective & 5 & 61.3 \\
ReDimNet-M-ft + one head, Demucs, AHC & 10 & 60.9 \\
\quad + averaged input, pick by objective & 10 & 63.5 \\
\rowcolor{rowhl}\quad + size prior ($\lambda=0.01$): \textbf{VoicePalette} & 10 & \textbf{63.9} \\
\midrule
Five such heads ensembled (not one model) & 5 & 64.7 \\
Best of the 52 candidates (oracle) & 10 & 72.8 \\
k-means from the true singer centroids$^b$ & --- & 78.7 \\
\bottomrule
\end{tabular}\\[2pt]
{\footnotesize $^a$Its encoder predates the dev set and may have seen dev songs. $^b$On the five-head ensemble's
embeddings.}
\end{table}

\Cref{tab:ladder} lists the steps on the development set. Three changes carry the gain from 59.3\% to 63.9\%. The
averaged lead/Demucs input (+0.8 points with AHC) uses the lead where it is clean and keeps the backing-vocal-free
Demucs evidence where the karaoke model is unsure. Choosing among k-means candidates by their own objective instead
of taking AHC (+1.2 to +2.9) exploits the head's geometry, which makes singers compact but not always separable by
average linkage. Replacing ECAPA2-ft with fine-tuned ReDimNet-M (+2.2 with the same decoding) is the largest step and
the only change of encoder that helped. Across the ten head seeds VoicePalette spans 62.7--65.3\%.

\subsection{Where the remaining error is}
\label{sec:ceiling}
The decomposition in \cref{tab:ladder} locates the error. In 72.8\% of lines the right answer is \emph{among} the 52
candidates, so a perfect chooser would add 9 points; k-means started from the true centroids reaches 78.7\%, so the
true partition is a stable fixed point of the geometry. Yet its objective is below that of the best restart in all
27 songs: the embedding space prefers wrong partitions, and no label-free criterion we tried recovers the right one
(\cref{sec:negative}). By pitch (Demucs input, AHC), accuracy falls from 68--69\% for lines whose median pitch is within two
semitones of the singer's median in the song to 45\% at 4--7 semitones and 39\% beyond: the main failure is the
same singer in a different register.

\subsection{Negative results}
\label{sec:negative}

\begin{figure}[t]
\centering
\includegraphics[width=\columnwidth]{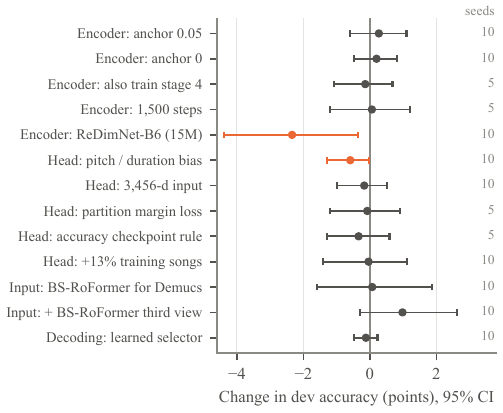}
\caption{Changes tested on VoicePalette after it was fixed (development set, difference in line accuracy to VoicePalette with the same seeds, paired artist-bootstrap 95\% CI). Orange: significantly worse. None is
significantly better.}
\label{fig:ablations}
\end{figure}

\Cref{fig:ablations} shows thirteen changes tested after VoicePalette was fixed, each with five or ten paired
seeds. \emph{Encoder:} the anchor weight (0, 0.05, 0.2), training one more stage, or three times as many steps give
the same accuracy, and the three-times-larger ReDimNet-B6 is significantly worse ($-2.3$ points). \emph{Head:} an
attention bias from pitch and duration differences is slightly worse; a learned input layer over the 3,456-d pooled
features that the encoder compresses to 192 dimensions ties, as do a decoder-aware partition margin loss, selecting
checkpoints by accuracy instead of \bthree{}, and 13\% more training songs. \emph{Input:} BS-RoFormer vocals in place of
Demucs tie; as a third view the gain (+1.0, CI $-0.3$ to $+2.6$) was not chosen by validation and is not claimed.
\emph{Decoding:} a selector trained on 187 songs unseen by encoder and head, over partition features (objective, size
entropy, agreement among candidates, centroid margin, share of pitch variance explained by the clusters, turn
persistence), picks the current rule in cross-validation and ties on dev.

Before VoicePalette was fixed, further ideas failed on the development set in the same way: short-line handling
(duration weighting, uncertainty weights, attaching short lines to long-line clusters), cross-register positives in
encoder training, pitch normalization of the input, a per-song context gate and singer slots in the head, a
frame-level same-singer verifier, fusion with a self-supervised singing model~\citep{torres2023singer}, a learned
co-membership decoder, and distilling the ensemble into one head (the five-head ensemble itself reaches only 64.7\%).

\paragraph{What the protocol caught.} Corpus validation ranked a learned view gate and a co-membership decoder above
the baseline; both lost on the development set. A five-seed mean moved by up to 0.9 points between seed sets
(anchor 0.05: +0.7 on seeds 2--6, $-0.1$ on seeds 7--11), and the selection heads of \cref{sec:negative} trained on
12\% fewer songs looked 1.1 points worse until the opposite test showed a tie. Any claim below about one point on this
development set needs ten seeds and a paired interval.

% ------------------------------------------------------------------------------------------------
\section{Limitations}
\label{sec:limits}
\begin{itemize}
  \item \textbf{Development-set size.} 27 songs resolve about 1.5 accuracy points; the development set was used for
  every decision and is optimistic for the final system.
  \item \textbf{Known singer count and line timings}, as in~\citep{ng2026linelevel}. Multi-singer lines are not part of the task.
  \item \textbf{Labels.} The 168 held-out-artist songs have single-annotator labels; both systems are scored on the
  same labels, which bounds absolute accuracy but not the comparison.
  \item \textbf{Separation} is outside the model and costs about 3.5\,min per song for the karaoke model on a laptop
  GPU. VoicePalette itself (encoder and head) runs in seconds per song.
  \item \textbf{Domain.} K-pop training data; J-pop session singers (IdolSongsJp) are the only other domain tested.
\end{itemize}

% ------------------------------------------------------------------------------------------------
\section{Conclusion}
VoicePalette, one song-context head on a fine-tuned ReDimNet-M encoder with an averaged lead/all-vocal input and candidate
selection by the k-means objective, improves line-level singer diarization on unseen artists by
$+0.033$ \bthree{} and 4.6 accuracy points over the ECAPA2-ft pipeline with sticky decoding and discography linking,
in one pre-registered test look, and matches a five-head ensemble. Thirteen further changes tie or lose under ten-seed paired intervals. The remaining error is not
in finding the right partition, which is among the candidates for 72.8\% of development lines, but in choosing it,
and in lines a singer sings outside their usual register; both need evidence beyond the voiceprint geometry, such as
lyrics, the song's structure or the music video.

\section*{Code and Models}
Code, the fine-tuned encoder and the ten song-context heads are available at
\url{https://github.com/nagiflow/voicepalette} (tag \texttt{paper-v1}); \texttt{reproduce.py} recomputes the
ColorCoded-SD result from audio.

\section*{Ethics and Data Statement}
ColorCoded is a web application operated by nagiflow LLC, the author's company; its annotations are user-created
projects. The service's terms of service (revised effective December 24, 2026) permit their use for research and the
publication of de-identified datasets derived from them, with an opt-out for users. User identifiers were hashed and no
personal data is used. The ColorCoded-SD benchmark is released as described in~\citep{ng2026linelevel}; no audio, lyric text or user
information is released with this paper. IdolSongsJp was used for non-commercial evaluation only, under its license;
no model was trained on it. Singer identification could be misused to track individuals; our task groups lines within
commercially released songs and does not identify anyone outside them.

\bibliography{refs}

\end{document}